\input harvmac
\input amssym.def
\input amssym.tex
\noblackbox
\newif\ifdraft

\catcode`\@=11
\newif\iffrontpage
\newif\ifxxx
\xxxtrue

\newif\ifad
\adtrue
\adfalse

\def\a{\alpha}
\def\b{\beta}

\def\r{r}

\def\dst{\displaystyle}
\def\gz{\dst\mathrel{\mathop g^{\scriptscriptstyle{(0)}}}{}\!\!\!} 
\def\go{\dst\mathrel{\mathop g^{\scriptscriptstyle{(1)}}}{}\!\!\!} 
\def\gt{\dst\mathrel{\mathop g^{\scriptscriptstyle{(2)}}}{}\!\!\!} 
\def\gn{\dst\mathrel{\mathop g^{\scriptscriptstyle{(n)}}}{}\!\!\!} 
\def\kz{\dst\mathrel{\mathop k^{\scriptscriptstyle{(0)}}}{}\!\!\!} 
\def\ko{\dst\mathrel{\mathop k^{\scriptscriptstyle{(1)}}}{}\!\!\!} 
\def\Rz{\dst\mathrel{\mathop R^{\scriptscriptstyle{(0)}}}{}\!\!} 
\def\Dz{\dst\mathrel{\mathop D^{\scriptscriptstyle{(0)}}}{}\!\!} 
\def\sqrtgz{\dst\mathrel{\mathop {\sqrt{g~}}^{\scriptscriptstyle{~(0)}}}{}\!\!\!} 
\def\sqrtgo{\dst\mathrel{\mathop {\sqrt{g~}}^{\scriptscriptstyle{~(1)}}}{}\!\!\!} 

\hfill

\def\{{\lbrace}
\def\}{\rbrace}

\def\a{\alpha}
\def\b{\beta}
\def\c{\gamma}

\def\frac#1#2{{\scriptstyle{#1}\over\scriptstyle{#2}}}

\noblackbox

\def\abstract#1{
\vskip.5in\vfil\centerline
{\bf Abstract}\penalty1000
{{\smallskip\ifx\answ\bigans\leftskip 2pc \rightskip 2pc
\else\leftskip 5pc \rightskip 5pc\fi
\noindent\abstractfont \baselineskip=12pt
{#1} \smallskip}}
\penalty-1000}

%
\lref\ACHU{A.~Achucarro and P.K.~Townsend ,'' A Chern-Simons action
 for three-dimensional anti-de Sitter supergravity theories ,``
  Phys.\ Lett.\ B {\bf 180} , 89 (1986) .}
%
%
\lref\Ali{
A.~H.~Chamseddine,
``Topological Gravity And Supergravity In Various Dimensions,''
Nucl.\ Phys.\ B {\bf 346}, 213 (1990).}
%
\lref\DS{
S.~Deser and A.~Schwimmer,
``Geometric Classification Of Conformal Anomalies In Arbitrary Dimensions,''
Phys.\ Lett.\ B {\bf 309}, 279 (1993)
[arXiv:hep-th/9302047].}
%
\lref\STone{
A.~Schwimmer and S.~Theisen,
``Diffeomorphisms, anomalies and the Fefferman-Graham ambiguity,''
JHEP {\bf 0008}, 032 (2000)
[arXiv:hep-th/0008082].}
%
\lref\STtwo{
A.~Schwimmer and S.~Theisen,
``Universal features of holographic anomalies,''
JHEP {\bf 0310}, 001 (2003)
[arXiv:hep-th/0309064].}
%
\lref\ISTY{
C.~Imbimbo, A.~Schwimmer, S.~Theisen and S.~Yankielowicz,
``Diffeomorphisms and holographic anomalies,''
Class.\ Quant.\ Grav.\  {\bf 17}, 1129 (2000)
[arXiv:hep-th/9910267].}
%
\lref\dHSS{
S.~de Haro, S.~N.~Solodukhin and K.~Skenderis,
``Holographic reconstruction of spacetime and renormalization in the  AdS/CFT
correspondence,''
Commun.\ Math.\ Phys.\  {\bf 217}, 595 (2001)
[arXiv:hep-th/0002230].}
%
\lref\WittenCS{
E.~Witten,
``Quantum Field Theory And The Jones Polynomial,''
Commun.\ Math.\ Phys.\  {\bf 121}, 351 (1989).}
%
\lref\MS{
G.~W.~Moore and N.~Seiberg,
``Taming The Conformal Zoo,''
Phys.\ Lett.\ B {\bf 220}, 422 (1989).}
\lref\Elitzur{
S.~Elitzur, G.~W.~Moore, A.~Schwimmer and N.~Seiberg,
``Remarks On The Canonical Quantization Of The Chern-Simons-Witten Theory,''
Nucl.\ Phys.\ B {\bf 326}, 108 (1989).}
%
\lref\Ramallo{
J.~M.~F.~Labastida and A.~V.~Ramallo,
``Operator Formalism For Chern-Simons Theories,''
Phys.\ Lett.\ B {\bf 227}, 92 (1989).}
%
\lref\BH{
J.~D.~Brown and M.~Henneaux,
``Central Charges In The Canonical Realization Of Asymptotic Symmetries: An
Commun.\ Math.\ Phys.\  {\bf 104}, 207 (1986).}
%
\lref\WittenGR{
E.~Witten,
``(2+1)-Dimensional Gravity As An Exactly Soluble System,''
Nucl.\ Phys.\ B {\bf 311}, 46 (1988).}
%
\lref\Maldacena{
J.~M.~Maldacena,
``The large N limit of superconformal field theories and supergravity,''
Adv.\ Theor.\ Math.\ Phys.\  {\bf 2}, 231 (1998)
[Int.\ J.\ Theor.\ Phys.\  {\bf 38}, 1113 (1999)]
[arXiv:hep-th/9711200].}
%
\lref\CHvD{
O.~Coussaert, M.~Henneaux and P.~van Driel,
``The Asymptotic dynamics of three-dimensional Einstein gravity with a negative
Class.\ Quant.\ Grav.\  {\bf 12}, 2961 (1995)
[arXiv:gr-qc/9506019].}
%
\lref\B{
M.~Ba\~nados,
``Global Charges In Chern-Simons Field Theory And The (2+1) Black Hole,''
Phys.\ Rev.\ D {\bf 52}, 5816 (1996)
[arXiv:hep-th/9405171].}
%
\lref\BTrZ{M.~Ba\~nados, R.~Troncoso and J.~Zanelli,
``Higher dimensional Chern-Simons supergravity,''
Phys.\ Rev.\ D {\bf 54}, 2605 (1996)
[arXiv:gr-qc/9601003].
R.~Troncoso and J.~Zanelli,
``New gauge supergravity in seven and eleven dimensions,''
Phys.\ Rev.\ D {\bf 58}, 101703 (1998)
[arXiv:hep-th/9710180].}
%
\lref\BGH{
M.~Ba\~nados, L.~J.~Garay and M.~Henneaux,
``The dynamical structure of higher dimensional Chern-Simons theory,''
Nucl.\ Phys.\ B {\bf 476}, 611 (1996)
[arXiv:hep-th/9605159].}
%
\lref\HS{M.~Henningson and K.~Skenderis,
``The holographic Weyl anomaly,''
JHEP {\bf 9807}, 023 (1998)
[arXiv:hep-th/9806087].}
%
\lref\TZ{C.Teitelboim and J.Zanelli, Class.Quant.Grav.{\bf 4}, L125 (1987)}
\lref\BTZ{M.~Ba\~nados, C.~Teitelboim and J.~Zanelli,
``Dimensionally Continued Black Holes,''
Phys.Rev.D {\bf 49}, 975 (1994)
[arXiv:gr-qc/9307033].}
\lref\KS{K.~Skenderis and S.~N.~Solodukhin,
``Quantum effective action from the AdS/CFT correspondence,''
Phys.\ Lett.\ B {\bf 472} (2000) 316
[arXiv:hep-th/9910023].}
%
\lref\OTZ{P.~Mora, R.~Olea, R.~Troncoso, and J.~Zanelli, in preparation.} 
\lref\BD{D.~G.~Boulware and S.~Deser,
``String Generated Gravity Models,''
Phys.\ Rev.\ Lett.\  {\bf 55}, 2656 (1985).}
%
\lref\BGN{
M.~Blau, K.~S.~Narain and E.~Gava,
``On subleading contributions to the AdS/CFT trace anomaly,''
JHEP {\bf 9909}, 018 (1999)
[arXiv:hep-th/9904179].}
%
\lref\NO{
S.~Nojiri and S.~D.~Odintsov,
``On the conformal anomaly from higher derivative gravity in AdS/CFT
correspondence,''
Int.\ J.\ Mod.\ Phys.\ A {\bf 15}, 413 (2000)
[arXiv:hep-th/9903033].}
%
\lref\Graham{R.~Graham, 
``Volume and Area Renormalizations for Conformally Compact Einstein Metrics,''
Rend. Circ. Mat. Palermo, Ser.II, Suppl. {\bf 63} (2000) 31 
[arXiv:math.DG/9909042].}
\lref\FG{C.~Fefferman and R.~Graham, 
``Conformal Invariants'', in {\it The mathematical heritage of Elie Cartan
(Lyon 1984)}, Ast\'erisque, 1985, Numero Hors Serie, 95.}
\lref\PS{I.~Papadimitriou and K.~Skenderis, 
``AdS/CFT correspondence and Geometry'', hep-th/0404176.}

\lref\MM{D.~Martelli and W.~Muck,
``Holographic renormalization and Ward identities with the  Hamilton-Jacobi
method,'' Nucl.\ Phys.\ B {\bf 654}, 248 (2003)
[arXiv:hep-th/0205061]; 
J.~Kalkkinen, D.~Martelli and W.~Muck,
``Holographic renormalisation and anomalies,''
JHEP {\bf 0104}, 036 (2001)
[arXiv:hep-th/0103111].}
%
%

\Title{\vbox{ \rightline{\vbox{\baselineskip12pt
\hbox{AEI-2004-034} \hbox{hep-th/0404245}}}}}
{{Chern-Simons Gravity and Holographic Anomalies}
\footnote{$^{\scriptscriptstyle*}$}{\sevenrm Partially
supported by GIF, the German-Israeli Foundation for Scientific
Research, Minerva Foundation, the Center for Basic Interactions 
of the Israeli Academy of Sciences and the Einstein Center.}}
\vskip 0.3cm
\centerline{M.~Ba\~nados$^a$, A.~Schwimmer$^b$ and S.~Theisen$^c$ } 
\vskip 0.6cm
\centerline{$^a$ \it Departament de F\'\i sica, 
P. Universidad Cat\'olica de Chile, 
Casilla 306, Santiago 22, Chile} 
\vskip.2cm
\centerline{$^b$ \it Department of Physics of Complex Systems,
Weizmann Institute, Rehovot 76100, Israel} 
\vskip.2cm
\centerline{$^c$ \it Max-Planck-Institut f\"ur Gravitationsphysik,
Albert-Einstein-Institut, 14476 Golm, Germany}

\abstract{We present a holographic treatment of 
Chern-Simons (CS) gravity theories in odd dimensions.  
We construct the associated holographic stress tensor and   
calculate the Weyl anomalies of the dual CFT.     
}

\Date{\vbox{\hbox{\sl {April 2004}}
}}
\goodbreak

\parskip=4pt plus 15pt minus 1pt
\baselineskip=15pt plus 2pt minus 1pt

\noblackbox

\newsec{Introduction and summary}

The  existence of a duality between gravitational theories in odd dimensions
and conformal field theories living on the boundary was first indicated
by the remarkable observation of Brown and Henneaux \BH\ that 
the asymptotic group of symmetries of 2+1 gravity with a
negative cosmological constant $\Lambda =-2/l^2$
is the two-dimensional conformal group with a non-vanishing central charge 
$$
c = {3l \over 2G}.
$$

Once it was understood that three-dimensional gravity can be
written as a Chern-Simons(CS) theory \ACHU,\WittenGR\ and 
that generically three dimensional CS theories are related to two dimensional
conformal field theories \WittenCS,  a more explicit relation underlying
the Brown-Henneaux argument became available \CHvD\ (see \B\ for a 
previous attempt).

With the arrival of Maldacena's conjecture \Maldacena, these results became special
cases of a much larger connection between gravitational and field theories.

The three dimensional CS gravity theory has some very special features
not having propagating degrees of freedom. Its generalization to
higher (odd) dimensions \Ali\ is a fully interacting theory which makes 
its study much more difficult. The corresponding supergravity theories were
formulated in \Ali\BTrZ\ and the Hamiltonian structure of the theory was
studied in \BGH.

In the present work we study a few basic issues related to a holographic
interpretation of higher dimensional CS theories. We construct the holographic
stress tensor and calculate the conformal anomalies, as first done in \HS\ for standard gravity,
of the (even dimensional) conformal field theory dual to the CS theory. 
The calculations have rather unusual features due to  two characteristic
properties of CS theories:

a) Even though the CS action when rewritten in metric form contains
higher powers of the curvatures, the equations of motion are polynomial
in the curvatures, containing at most second derivatives of the metric.

b) The AdS solution is  $n$-th degenerate in $d=2n+1$ dimensions and therefore
the expansion around the solution starts with a $n+1$-th order term.

Due to property b) the standard Fefferman-Graham expansion \FG\
breaks down  and the usual methods of evaluating the conformal anomalies
cannot be used. We calculate the anomalies in three different ways:
 
i) We use the coefficients of the CS action in integer dimension in
dimensional regularization when property b) is not obeyed and then we take
the limit to integer dimension. The values of the anomalies are obtained
from the general formulae for actions with higher powers of the curvatures
\NO\BGN\STtwo. 
 
ii) We use a dimensionally continued CS action where the dimension 
dependent coefficients are tuned in order that a), b) are obeyed and
we calculate the anomalies from the equations of motion.
   
iii) In integer dimensions, we derive the stress tensor from the equations 
of motion, and find a general formula valid for all $n$. We also calculate 
the stress tensor in the Hamiltonian formalism. 

\noindent
For i) and ii) we only present results for $n=2$, but for iii) 
we give explicit results for any $n$. 

All the different ways of calculating the anomalies  agree and give the 
results: 
 
a) All type B anomalies vanish (we remind that their number increases 
with the dimension)

b) The type A anomalies (one in each even dimension)
are nonzero and  consistent with the universal formula of \ISTY\ for the specific CS action. 

Feature a) restricts a direct holographic interpretation of the CS action.
Through the diffeo Ward identities the type B anomalies are related
to lower order correlators of energy momentum tensors and one of them to the 
two point function. Its vanishing cannot happen in a unitary conformal theory.
On the other hand b) shows that the CS gravitational actions provide
the correct analogue of the CS gauge actions which generate
the chiral anomalies for generating the type A trace anomalies.

The outline of the paper is as follows. In the next section we review 
the features of  CS gravity which are relevant for the 
further developments. In sect.~3 we present the holographic analysis 
in the spirit of refs.\ISTY\STone\STtwo. For this we need to 
define a dimensionally continued CS gravity action. We do this 
explicitly for the case appropriate to a four-dimensional boundary 
theory. 
In integer dimensions the powerful method of differential forms, in which the 
CS theory is most naturally formulated, leads to a very efficient 
method, based on the equations of motion,  
to derive the holographic stress tensor for Chern-Simons gravity 
in any odd dimension. This is done in sect.~4.
In sect.~5 we present a Hamiltonian derivation 
of the stress tensor which is applicable for any gravity theory
(cf. also \MM\PS). 
As a first demonstration of this method we reproduce 
the known result for ordinary gravity. Next we apply it to CS gravity which
turns our to be much simpler. As a final application of our formula for the 
stress tensor we use it to compute the mass of CS black holes. 


\newsec{Review of Chern-Simons gravity} 

The action of euclidean CS gravity in $2n+1$ dimensions is \Ali
\eqn\CSa{
I_{2n+1}=\int_{M_{2n+1}}\omega_{2n+1}\,.}
$\omega_{2n+1}$ is a CS $(2n+1)$-form for the group 
$SO(1,2n+1)$, i.e. 
\eqn\CSform{
d\omega_{2n+1}=F^{A_1 A_2}\wedge \cdots\wedge F^{A_{2n+1}A_{2n+2}}
\epsilon_{A_1\dots A_{2n+2}}}
where $F^{AB}$ is the curvature two-form. 
This action is invariant under
gauge transformations, up to a boundary term.

To exhibit the gravitational character of the CS action one
splits the $SO(1,2n+1)$ 
indices $A=(0,a)$ and decomposes the gauge potential according to
\foot{Hatted quantities are defined in the $(2n+1)$-dimensional 
bulk. Unhatted symbols will be used to  below  
and refer to quantities defined on the $2n$-dimensional boundary.}
\eqn\Adef{
A={1\over2}A^{AB}J_{AB}={1\over2}\hat\omega^{ab}J_{ab}+\hat e^a P_a}
where 
\eqn\AJdef{
A^{0a}=\hat e^a\,,\qquad J_{0a}=P_a\,,}
\eqn\Fieldstrength{
F={1\over2}F^{AB}J_{AB}={1\over2}(\hat R^{ab}+\hat e^a \hat e^b)J_{ab}+\hat T^a P_a}
and
\eqn\RTdef{\eqalign{
\hat R^{ab}&=d\hat\omega^{ab}+\hat\omega^a{}_c\hat\omega^{cb}\,,\cr
\hat T^a&=d\hat e^a+\hat\omega^a{}_b \hat e^b\,.}}
Using this decomposition the action becomes, up to a boundary term,  
\eqn\CSb{\eqalign{
I_{2n+1}&={\int_{M_{2n+1}}}\!\!\!\!\!\!
\epsilon_{\a_1\dots \a_{2n+1}}
\sum_{p=0}^n{1\over 2(n-p)+1}{n \choose p}
\hat R^{\a_1 \a_2}\wedge\cdots\wedge\hat R^{\a_{2p-1}\a_{2p}}
\wedge\hat e^{\a_{2p+1}}\wedge\dots\wedge\hat e^{\a_{2n+1}}\cr
&=\int_{M_{2n+1}}\!\!\!\!\!\!\!\!\!d^{2n+1}x \sqrt{\hat g}~ 
\sum_{p=0}^n{n\choose p}[2(n-p)]!\,\hat E_{2p}\,.}}
In \CSb\ the expressions $\hat E_{2n}$ are the Euler densities 
\eqn\EulerNorm{
\eqalign{E_{2n}&\equiv {1\over 2^n}R_{i_1 j_1 k_1 l_1}
\dots R_{i_n j_n k_n l_n}\epsilon^{i_1 j_1 i_2 j_2\dots i_n j_n}
\epsilon^{k_1 l_1 k_2 l_2\dots k_n l_n}\cr
&=R^n+\dots}}
The integral of $\int_M\sqrt{g}\, E_{2n}$ over a $(2n)$-dimensional manifold 
without boundary is a topological invariant of $M$. 
While the expression in the first line is only meaningful in 
$2n$ dimensions, the expression in the second line can be defined in 
any dimension. For instance, for $n=2$ one finds explicitly
\eqn\Eulerfour{
E_{4}={1\over4}R_{ijkl}R_{mnpq}\epsilon^{ijmn}\epsilon^{klpq}
=R^2-4 R^{ij}R_{ij}+R_{ijkl}R^{ijkl}\,.}
For many purposes the form of the action as written in the first line 
of \CSb\ is more convenient, in particular for formal manipulations. 
However it is the second form which allows continuation to
non-integer dimensions. 

If one allows in \CSb\ for arbitrary relative coefficients for the 
$\hat E_{2p}$ one arrives at Lovelock gravity. It is, however, 
only for the special coefficients, namely those of CS-gravity, 
that the manifest $SO(2n+1)$ symmetry is enhanced to $SO(1,2n+1)$. 
This makes CS gravity in many respects very `non-generic', as we 
will see in the following.   

{}From \CSb\ one derives the following equation of motion for the vielbein $\hat e_\mu^a$:
\eqn\eomvielbein{
\epsilon_{\mu_1\dots\mu_{2n+1}}F^{\mu_1\mu_2}\wedge\cdots\wedge F^{\mu_{2n-1}\mu_{2n}}=0}
where 
\eqn\Fmunu{
F^{\mu\nu}=\hat R^{\mu\nu}+dx^\mu\wedge dx^\nu.}
The equation of motion for the spin connection $\omega$
is solved imposing the torsion constraint $\hat T=0$. 
This is the general solution for $n=1$. In more than three dimensions 
there are other solutions. We will, however, always impose $\hat T=0$ and 
thus the only degrees of freedom are those of the metric. 

As it is clear from \Fmunu\ the vanishing of $F^{\mu\nu}$ is equivalent
to the $2n+1$ dimensional metric being AdS. Then \eomvielbein\ shows
that after the torsion constraint is taken into account AdS is an $n$ fold
degenerate solution of the equations of motion. As a consequence an expansion
around the AdS solution will start with $n+1$ order terms.

\newsec{Dimensionally continued Chern-Simons gravity}

In this section we will restrict ourselves to an action with at most 
four-derivatives. This is appropriate for the dimensional continuation 
of five-dimensional CS gravity. The most general such action is 
\eqn\quartic{
S=\int d^{d+1}x\sqrt{G}\left(\hat R-2\Lambda+\a \hat R^2
+\b \hat R_{\mu\nu}\hat R^{\mu\nu}
+\c \hat R_{\mu\nu\rho\sigma}\hat R^{\mu\nu\rho\sigma}\right)\,.}
One needs\foot{We use the following sign convention for the Riemann tensor:
$[\nabla_\mu,\nabla_\nu]V_\rho=R_{\mu\nu\rho}{}^\sigma V_\sigma$.}
$\Lambda=-{1\over2}d(d-1)+{1\over2}(d-3)\left(\a d^2(d+1)+\b d^2+2\c d\right)<0$ 
for $AdS_{d+1}$ to be a solution of the equations of motion. 
This is required for the AdS/CFT correspondence.  
The Weyl anomaly for the dual conformal field theory in $d=4$ was computed in
\NO\BGN\STtwo\ with the result
\eqn\quarticanomaly{
\langle T^i_i\rangle={1\over8}\left\lbrace
(1-40\a-8\b+4\c)C^2-(1-40\a-8\b-4\c)E_4\right\rbrace\,.}
For $d=4$, $\b=-4\a=-4\c=-1$ and $\Lambda=-3$ \quartic\ becomes the 
action of CS gravity in five dimensions. For these values of the parameters
the anomaly \quarticanomaly\ is purely type A. However, it is \`a 
priori not obvious that this result is reliable since it was obtained 
under the condition that the generic FG-expansion is valid.  

We have mentioned in the 
introduction the two special features of this action, namely that 
the equations of motion contain no higher than second derivatives of the 
metric and that when expanded around $AdS_5$ the expansion starts at 
cubic order. The first feature is maintained as long as $\b=-4\a=-4\c$. 
This means that at each order in curvatures they appear in 
the Euler combination. The second feature requires
\eqn\CSparameters{
\Lambda=-{1\over4}d(d-1)\,,\qquad\a=\c=-{1\over4}\b={1\over2(d-2)(d-3)}\,}
as a tedious calculation reveals. The action \quartic\ with the choice 
\CSparameters, which is the dimensionally continued 
five-dimensional CS action, will be the starting point of our analysis. 

We make the FG expansion for the bulk metric $G_{\mu\nu}$, i.e. 
we make the Ansatz \FG
\eqn\FGAnsatz{
ds^2={1\over4}\left({dr\over r}\right)^2
+{1\over r}g_{ij}(x,r)dx^i dx^j}
with
\eqn\gexpansion{
g_{ij}(x,r)=\sum_{n=0}^\infty \gn_{ij}(x)r^n\,.}
The coefficients $\gn_{ij}$ are to a large extent fixed by the 
symmetries, i.e. invariance of \FGAnsatz\ under so-called PBH transformations \ISTY. 
To fix them completely one inserts the Ansatz into the equations of motion
and solves for the $\gn_{ij}$ recursively. They can be expressed in terms of 
curvature tensors constructed from $\gz_{ij}$ with a total of 
$2n$ derivatives of the metric. For generic gravitational actions, e.g. 
\quartic, the $\gn_{ij}$ are all local as long as one stays away from 
integer dimensions. For instance, the PBH transformations completely 
fix the local part of $\go_{ij}$ to 
\eqn\gonelocal{
\go_{ij}^{\rm~loc}=-{1\over(d-2)}\left(\Rz_{ij}-{1\over2(d-1)}\Rz\gz_{ij}\right)\,.}
The only freedom left is an additive term which is
a tensor built from $\gz_{ij}$ which is invariant under Weyl 
transformations of $\gz_{ij}$ and transforms homogeneously of order
$-2$ under a constant rescaling of the coordinates. Clearly this 
cannot be a finite polynomial in the curvature tensors. Terms containing
tensors constructed from $C_{ijkl}$, the Weyl tensor, multiplied 
with powers of $\sqrt{C^2}$ have the right transformation properties . 
We will see that such a term is required in CS gravity.

Before writing down the equations of motion we want 
to compute the Weyl anomaly for CS-gravity for $d=4$. 
{}From \ISTY\ we know that it is given (in $d=2n$) 
by the ${\cal O}(\r^{-1})$ coefficient of the expansion of the 
gravitational action. For CS gravity and $n=2$ one finds
\eqn\anomalyCS{
\langle T^i_i\rangle_{CS}={1\over 4}E_4\,,}
which agrees with what we found before. While the previous derivation 
used the explicit form for $\go_{ij}$ this derivation 
does not need the explicit form of any of the coefficients 
$\gn$. This is a consequence of the choice \CSparameters. 

We now return to the equations of motion. 
For \quartic\ with the choice \CSparameters\ one finds that 
at lowest orders in $r$ the equations of motion are identically satisfied. 
The first non-trivial equations are at ${\cal O}(\r^0)$ for the 
$(\r\r)$ components and at ${\cal O}(\r)$ for the 
$(ij)$ and $(i\r)$ components. At these orders the $\gt$ 
dependence drops out identically as a consequence of the choice 
\CSparameters.

The equations can now be written as the definition of 
the non-local conserved energy-momentum tensor 
$T_{ij}$, its trace and conservation. One finds 
\eqn\eom{\eqalign{
T_{ij}&={1\over24(d-4)}\left(C_{iklm}C_j{}^{klm}-{1\over4}C^2 g_{ij}\right)
+\hbox{finite as $d\to 4$}\,,\cr
\nabla^i T_{ij}&=0\,,\cr
g^{ij}T_{ij}&=-{1\over4}(R^2-4 R_{ij}R^{ij}+R_{ijkl}R^{ijkl})}}
where $T_{ij}$ is the following expression in terms of $\go_{ij}$:
\eqn\defT{\eqalign{
&T_{ij}={1\over6}\left\lbrace 
(d-3)\Bigl(2 \go^{~2}_{ij}-2 \tr(\go)\,\go_{ij}+(\tr\go)^2 g_{ij}
-\tr(\go^{~2})g_{ij}\Bigr)\right.\cr
\noalign{\vskip.2cm}
&\left.+2(R_i{}^k\go_{jk}+R_j{}^k\go_{ik})
+2R_{ikjl}\go^{~kl}+R\,\tr(\go)g_{ij}-2R_{kl}\go^{~kl}\,g_{ij}
-R\go_{ij}-2\,\tr(\go)R_{ij}\right\rbrace\,.}}
In these expressions all curvature tensors are computed with $\gz$
which is also used to raise indices, 
$g_{ij}$ stands for $\gz_{ij}$ and 
$C$ is they Weyl tensor which is totally traceless in 
$d$-dimensions.\foot{On the r.h.s. of the first of eq.\eom\ the 
finite piece, which is traceless in $d=4$, 
is absent if one instead interprets $C$ as the 
Weyl-tensor which is traceless in $d=4$ but with the range of its 
indices extended to $d$.}
The first term on the r.h.s. of eq.\eom\ becomes ${0\over0}$ in 
$d=4$ due to a special identity. 

Eq.\eom$_1$ together with \defT\ determine $\go$. We have not obtained a 
closed expression for $\go$ itself but clearly it cannot be local 
in $d=4$ (see also the discussion in \STtwo). 
The unique local expression 
for $\go_{ij}$ which solves the PBH equation has to be augmented by a 
non-local piece which is invariant under Weyl transformations. 
We write 
\eqn\gone{
\go_{ij}=\go_{ij}^{\rm~loc}+\Delta_{ij}\,.}
Using this definition, the $(ij)$ component of the equation motion at 
${\cal O}(\r)$ can be written in the form
\eqn\eqnij{\eqalign{
C^{(d)}_{ikjl}\Delta^{kl}&+{1\over2}(d-3)\left\lbrace 2\Delta^2_{ij}
-2(\tr\Delta)\Delta_{ij}+(\tr\Delta)^2\gz_{ij}-\tr(\Delta^2)\,\gz_{ij}\right\rbrace\cr
&={1\over8(d-4)}\left\lbrace C^{(d)2}_{ij}-{1\over4}C^{(d)2}\gz_{ij}\right\rbrace\,.}}
What we found here is reminiscent of the situation for the generic gravitational action:
there the equations of motion determine $\gt_{ij}$ which is local but has a pole 
at $d=4$. However,  
$\gt_{ij}$ consists of two cohomologically non-trivial pieces: one term which 
is the same (up to an overall coefficient) as the r.h.s. of \eqnij\ and 
an other which has a genuine pole (i.e. finite residue) 
at $d=4$. They are related to 
type A and type B anomalies. Here the cohomologically non-trivial information
resides in the expression on the r.h.s. of \eqnij. The significance of the 
particular combination of non-local terms $\Delta_{ij}$ 
is not clear to us other that it produces a local expression. 

We remark that in a holographic interpretation the Conformal Field Theory
living on the boundary will be necessarily non-unitary .
This is a consequence of the vanishing of the type B anomaly .
As it is well known this anomaly can be related to the correlator of two energy momentum 
tensors which cannot vanish in a unitary theory .  

\newsec{Chern-Simons stress tensor. Integer dimensions analysis }

In this and the following section we will rederive the results of sect.~3 using the equations of motion and 
action in integer dimensions. We use in this section the following form of the Chern-Simons equations of 
motion (c.f. \eomvielbein)
\eqn\CSeq{
\varepsilon_{\mu\nu\lambda\rho\sigma} (\hat R^{\mu\nu} +
dx^\mu \wedge dx^\nu)\wedge(\hat R^{\lambda\rho} + dx^\lambda \wedge dx^\rho)=0}
where 
\eqn\Rie{ \hat R^{\mu\nu} = {1 \over 2}  \hat R^{\mu\nu}_{~~ \lambda\rho} dx^\lambda \wedge dx^\rho} 
is the 2-form Riemann tensor. We shall see that this notation provides a powerful way to identify the 
stress-tensor in Chern-Simons gravity. 

This section is organized as follows. We first review some standard results and continue by making 
the connection with the FG expansion \FGAnsatz\ and \gexpansion. We then treat the 
five-dimensional case and recover the results of the previous section. 
We then apply the formalism to three and seven dimensional CS gravity, and finally provide a general 
formula for the Chern-Simons holographic stress tensor valid in an arbitrary dimension $D=2n+1$.  

\subsec{The FG expansion and Gauss-Codazzi equations}

Consider the space-time metric in normal coordinates
\eqn\metricH{
ds^2 = N^2(r) dr^2 + h_{ij}(r,x^i) dx^i dx^j}
and introduce the standard notation
\eqn\excurv{
K_{ij} = -{1 \over 2N} h_{ij}'} 
where the prime denotes derivatives w.r.t. $r$. 
The space-time curvature can be decomposed in the Gauss-Codazzi form 
\eqn\GaussCodazi{\eqalign{
&  \hat R^{ir}_{\ \  kl} = {1 \over N} (K^i_{\ k/l} - K^i_{\ l/k})\,, \cr
&  \hat R^{ir}_{\ \  jr} = {1 \over N} K^{i\prime}_{\ j} - K^i_{\ l} K^l_{\ j}\,, \cr
&  \hat R^{ij}_{\ \  kl} =  R^{ij}_{\ \  kl} - K^i_{\ k} K^j_{\ l} + K^i_{\ l} K^j_{\ k}}}
where $/$ represents the $2n$-dimensional covariant derivative in the metric 
$h_{ij}$.  Introducing the curvature 2-form $R^{ij}$ and extrinsic curvature 1-form $K^i = K^i_{\ j} dx^j$, 
these expressions can be rewritten more compactly as
\eqn\curvs{\eqalign{
& \hat R^{ir}=-{1\over N} D K^i + \left({1\over N}K^{i '} - K^i_{\ j} K^j\right)\wedge dr\,, \cr
& \hat R^{ij}=R^{ij}-K^i \wedge K^j + N(K^{i/j}-K^{j/i})\wedge dr\,.}}
We now make contact with \FGAnsatz\ by making a definite choice of the radial coordinate, i.e.,  
\eqn\Ansatzshift{
N = {1 \over 2r}}
and introduce the metric $g_{ij}$ as
\eqn\metrich{ 
h_{ij}(r,x^i) = {1 \over r} g_{ij} (r,x^i)\,.}
Then, it follows,
\eqn\Kijdd{
K_{ij} = {1 \over r} g_{ij} - g_{ij}' \ \ \ \Rightarrow \ \ \  K^i
= dx^i -r k^i}
where we have defined $k_{ij}=g_{ij}'$ and $k^i_{\ j} = g^{ik} g_{kj}' \,.$

Since the Christoffel symbols are invariant under constant rescaling of the metric, 
and multiplying by $r$ is a constant rescaling, the covariant derivatives  are not 
altered by the field redefinition $h_{ij} \rightarrow g_{ij}$.  

Recall now the definition of the $SO(1,5)$ curvature which enters in the CS equations of motion \CSeq\foot{Strictly speaking, $F^{\mu\nu}$ is the $SO(5)$ projection of the $SO(1,5)$ curvature (c.f. \Fieldstrength). } %
\eqn\SOcurvature{
F^{\mu\nu} = \hat R^{\mu\nu} +  dx^\mu\wedge dx^\nu\,.}
By direct computation we find
\eqn\curvcomp{\eqalign{
&  F^{ir} =   r^2 [2 Dk^i - (2k^{i\prime} + k^i_{\ j} k^j)\wedge dr ] \,, \cr
&  F^{ij} =   r \left[R^{ij} +  dx^i \wedge k^j + k^i \wedge dx^j - r
k^i \wedge k^j\right] + (`` ~ " ) \wedge dr \,.}}
(The components along  $dx^i\wedge dr$ in the second line will not be needed.)

So far we have not made any approximations. We can now use \gexpansion\ from where we derive
\foot{In EH gravity for $D$ odd the FG expansion \gexpansion\ needs to 
be modified. Starting at ${\cal O}(\rho^n)$ $\log(\rho)$ terms appear. Without the 
logarithmic terms the equations of motion for the $\gn$ are inconsistent. Working in 
non-integer dimensions, as we did in sect.~3, does not require the log terms for any 
gravitational theory. For CS gravity in integer dimensions, at least    
to the order we are considering here, they do not seem to be necessary either.} 
\eqn\gddexp{\eqalign{
 k_{ij}  &= \go_{ij} + 2r  \gt_{ij} +\cdots \,,\cr
 g^{ij}  &= \gz^{~ij} - r \go^{~ij} + r^2(-\gt^{~ij} + (\go^{~2})^{ij}) + \cdots \,,\cr
 k^i{}_j &= \go^{~i}{}_j+ r\left(2 \gt^{~i}{}_j-(\go^{~2})^{i}{}_j\right)+\cdots\,.}}
On the right hand side of these expressions, indices are lowered and 
raised with $\gz$.

\subsec{The Chern-Simons holographic stress tensor} 

Consider the Chern-Simons equations of motion \eomvielbein.
Our aim is to rederive the holographic stress tensor found in 
Sec. 3 using these equations. We shall first give an argument based only on the structure of 
these equations. In the next paragraph we prove, using a Hamiltonian approach, that our formula is in fact 
the variation of the renormalized action with respect to the boundary metric.  

We start for illustrative purposes with the five-dimensional case, but we shall see 
that for Chern-Simons theories of gravity the holographic stress tensor calculation is the same in all (odd) dimensions. We shall in fact provide a formula for this tensor valid on any (integer) dimension $d$.     
      
\noindent
{\it Five dimensional CS gravity.} The equations of motion were given in \CSeq. Let us study them in the 
lowest non-trivial order, i.e., the equations involving $\gz$ and $\go$. These are
\eqn\degeqcomp{\eqalign{
\varepsilon_{ijkl} F^{ij} \wedge F^{kl} = 0 &  \ \ \   \Rightarrow \ \
\  r^2\varepsilon_{ijkl}(\Rz^{\, ij} + 2 dx^i \wedge\go^{~j}) \wedge(\Rz^{\, kl} 
+ 2 dx^k \wedge \go^{~l})= 0 \,, \cr
4 \,\varepsilon_{ijkl} F^{ij} \wedge F^{kr} = 0 & \ \ \   \Rightarrow \ \
\  4\,r^3\varepsilon_{ijkl}(\Rz^{\, ij} + 2\ dx^i \wedge\go^{~j}) \wedge\Dz\go^{~k} =0}}
where, in the second column, we have rewritten the equations using \curvcomp\ and 
have kept only the lowest order terms.

Thanks to the Bianchi identity $\Dz\wedge \Rz^{ij}=0$ the
covariant derivative in the second equation can be pulled out to obtain
\eqn\Bianchi{
\Dz\,\Bigl(4\varepsilon_{ijkl} (\Rz^{\,ij} +  dx^i \wedge \go^{\,j})
\wedge \go^{~k}\Bigr) =0\,.}
We write this equation in the form
\eqn\rewrite{
4\varepsilon_{ijkl}(\Rz^{\,ij}+dx^i\wedge\go^{\,j})\wedge\go^{~k}=\tilde T_l}
where $\tilde T_l$ is an ``integration constant" 3--form that must be 
conserved $D\wedge
\tilde T_l=0$. Eq. \rewrite\ is an algebraic equation for $\go^{~i}$ which, in 
principle, can be
solved, and the solution involves the conserved tensor $T_{ij}$. The 
index structure of $\tilde T$ is $\tilde T^i_{\ npq}$ being antisymmetric in $npq$. We 
dualize and define a rank two tensor
\eqn\Tdual{
T^{ij} = {1 \over 3!}\varepsilon^{i npq} \tilde T^j_{\ \ npq}\,,}
which is symmetric
thanks to $\go_{ij}=\go_{ji}$ and $\Rz_{ijkl} = \Rz_{klij}$. 
Clearly, the conservation equation $D\wedge \tilde T^i=0$ in terms of $T^{ij}$ reads 
$ T^{ij}_{\  ~ ;j}=0$.

The conserved tensor is not completely arbitrary. In fact, the remaining equation of motion \degeqcomp\ fixes 
its trace $g_{ij}T^{ij}$ to be
equal to the four-dimensional Euler density. To see this we first 
 note that the trace $g_{ij} T^{ij}$ can also be expressed in terms of the 3-form $\tilde T_l$ in a convenient 
way. Since $^*\! (dx^i\wedge dx^j\wedge dx^k\wedge dx^l) =\epsilon^{ijkl}$ it follows
\eqn\TT{ g_{ij} T^{ij} =~ ^*\!\!\left( dx^i \wedge \tilde T_i \right).}
Hence, from the definition of $\tilde T_i$ (c.f. \rewrite) and \degeqcomp\ we find
\eqn\TE{
dx^l \wedge \tilde T_l = -4\varepsilon_{ijkl} (\Rz^{\,ij} +  dx^i
\wedge \go^{~j}) \wedge \go^{~k} \wedge dx^l =
\varepsilon_{ijkl} \Rz^{ij} \wedge \Rz^{kl}\,}
or, what is equivalent thanks to \TT, $g_{ij}T^{ij} = E_4$.

Of course, what we found here is just the set of equations \eom$_{2,3}$ and \defT\ evaluated at $d=4$.  
In fact, this structure is present for all Chern-Simons theories: the ambiguity in the FG-expansion always 
occurs in $\go$ and it equals the energy-momentum tensor $T_{ij}$. 

\noindent{\it A general formula.}  The analysis of the Chern-Simons equations in other integer dimensions reveals that the same structure appears in all cases. In $2n+1$ dimensions one finds   
\eqn\general{\tilde T_i = \int_0^1 dt\ 2n \,  \epsilon_{i_1 i_2...i_{2n-1}i} \left[ \Rz^{i_1 i_2} 
+ 2t  \ dx^{i_1} \wedge\go^{~ i_2} \right]^{n-1} \wedge \go^{~i_{2n-1}},  }
as can be checked for lower $n$ cases, and we have, in particular, checked $n\leq 3$. In \general, $t$ is an auxiliary parameter and the symbol $[~]^{n-1}$ means $n-1$ factors of the tensor $\Rz^{ij}+ 2t  \ dx^{i} \wedge\go^{~ j}$ contracting $2n-2$ indices in the Levi-Cevita symbol.  We now check in general that \general\ is conserved and its trace equal to $E_{2n}$, as a consequence of the $2n+1$ Chern-Simons equations of motion. 

Taking the covariant derivative of $\tilde T_i$ the integral over $t$ drops out and we find
\eqn\CDT{  D\tilde T_i =  2n\,\epsilon_{ i_1 i_2...i_{2n-1}i} \left[\Rz^{i_1 i_2} + 2 dx^{i_1}
\wedge \go^{~i_2} \right]^{n-2}  \wedge D\go^{~i_{2n-1}}  }
which is zero thanks to the $2n+1$ Chern-Simons equations.   In the same way, using the identity 
$(a+b)^n = a^n + n\int dt(a+t b)^{n-1}b$ and the equation of motion 
$\epsilon_{i_1...i_{2n}}\left[ \Rz^{i_1 i_2} + 2  \ dx^{i_1} \wedge\go^{~ i_2} \right]^{n}=0$, 
we can compute the trace $dx^i\wedge \tilde T_i$ and find the 2n-dimensional Euler density,
\eqn\tracen{ dx^i\wedge\tilde T_i=\epsilon_{i_1...i_{2n}} \Rz^{i_1 i_2}\wedge\cdots\wedge \Rz^{i_{2n-1} i_{2n}},}
as expected.

\newsec{Hamiltonian method}

We have found in the previous sections a general formula for the stress-tensor for Chern-Simons gravity, via dimensional regularization methods, and by a direct use of the equations of motion in integer dimensions. In this section we would like to rederive this formula as the functional variation of the effective action with respect to the boundary metric, 
\eqn\defTgeneral{
T_{ij} = {2 \over \sqrt{g_{(0)}}}
{\delta  I[g_{(0)}] \over \delta g_{(0)}^{ij}}\,.}
In the AdS/CFT correspondence, the functional $I$ is the regularized and renormalized 
bulk gravitational action written as a function of $g_{(0)}$. 

We shall be interested directly in the stress tensor and not in the effective action. 
The Hamiltonian formalism of gravity provides a shorter method to compute $T_{ij}$ which 
will be particularly convenient in Chern-Simons gravity. 

The computation of holographic anomalies via hamiltonian methods has also been considered 
in \MM\ and \PS. We shall briefly discuss the general idea and then apply it to Chern-Simons gravity. 

\subsec{The method}

If the metric is put in the ADM form
\eqn\ADMform{
ds^2 = N^2 dr^2 + h_{ij} (dx^i + N^i dr)(dx^j + N^j dr)}
the variation of the ADM action, evaluated on any solution of the 
equations of motion is
\eqn\dI{
\delta I = \int_{r=0} d^{2n}x\  \pi^{ij} \delta h_{ij}.}
In Einstein gravity,
\eqn\piEH{
\pi^{ij} = \sqrt{h} (K^{ij} - K h^{ij}) , \ \ \ \ \   K_{ij} = -{1 
\over 2N} h_{ij}'\,.}
In Chern-Simons gravity \dI\ will still be true, although the relation between the momenta  and extrinsic 
curvature will change. The formula \dI\ gives the variation of the action directly in  terms of a boundary 
integral evaluated at $r=0$. However, there are two problems with this  expression. First, it diverges and 
needs to be regularized and renormalized. Second, in FG  what is fixed is $\gz_{ij}$, not $h_{ij}$.

The first issue can easily be solved by adding covariant counterterms. We first 
regularize by evaluating at some fixed finite $r$.  The subtraction will be quite
straightforward.  The second problem is more delicate, but has a nice solution. We would 
like  the variation of the action to have the form 
$ \int T_{(\rm reg)}^{ij} \delta g_{(0)ij}$. However, replacing in \dI\
\eqn\hexpansion{
h_{ij} = {1 \over r}\left(\gz_{ij} + r \go_{ij} + r^2 \gt_{ij} + \dots \right)}
we get
\eqn\dIexp{
\delta I =  \int_{r} d^{2n}x  \left( {1 \over r} \pi^{ij}\,
\delta\!\!\gz_{ij}+ \pi^{ij}\, \delta\!\! \go_{ij}+ \cdots \right)\,.}
Now, the point is that the extra terms, $\int \pi^{ij}\,\delta\!\!\go_{ij}\dots$, 
can be transformed into contributions of the form 
$ \int A^{ij}\,\delta\!\!\gz_{ij}$ by making appropriated ``integral by parts", 
and discarding total variations.

Our prescription is then to expand \dIexp\ and make the necessary ``integral by parts"
until it has the form $\int T_{(\rm reg)}^{ij}\, \delta\!\!\gz_{ij}$, plus 
total 
variations.
Then, we discard all total variations, identify $T^{ij}_{(\rm reg)}$, 
renormalize by subtracting the divergent terms, and find $T^{ij}_{(\rm ren)}$.

The terms which are total variations, $\delta f(\gz,\go,\gt,\dots)$, must be discarded 
because they cannot be written, by means of integrals by parts, 
as $A^{ij}\,\delta\!\!\gz$. Hence, the Dirichlet problem 
dictates that we add to the action a boundary term $-f$ to cancel this 
variation.  

As a warm-up we will first treat the standard Einstein action.   
We should and will recover the energy momentum tensor found in \dHSS.  
\foot{We do not include the log terms.
Including then would simply mean a finite renormalization and it does 
not affect the trace of $T_{ij}$.} 
{}From \piEH\ and \Kijdd\ we get
\eqn\dIEH{
\delta I = \int \sqrt{g} \left[ - {1 \over r^{d/2-1} } (k^{ij}-k
g^{ij}) +  {1-d \over r^{d/2}} g^{ij}\right] \delta g_{ij} =- \int
{\sqrt{g} \over r^{d/2-1}}  (k^{ij}-k g^{ij})  \delta g_{ij}}
where the second term has been discarded because 
$\sqrt{g} g^{ij}\delta g_{ij} =2\delta \sqrt{g}$ is a total variation.

Consider first $d+1=3$. Since $r^{d/2-1}=1$ in this case, the variation of the action $\delta I$ is finite. 
And since $k_{ij} = \go_{ij} + ...$, its non-zero part is,
\eqn\dIthree{
\delta I =-\int\sqrt{\gz~}\,\Bigl(\go^{~ij}-\Tr(\go)\gz^{~ij}\Bigr)\,\delta\!\!\gz_{ij}}
giving the correct expression for $T^{ij}$ (see Eq. (3.10) in \dHSS).

Consider now $d+1=5$. In this case there is a divergent piece that we cancel by a 
subtraction. We focus on the finite piece obtained by expanding 
$\sqrt{g}(k^{ij}-k g^{ij})\delta g_{ij}$ to ${\cal O}(r)$. 
It is useful to note that $\sqrt{g} k g^{ij}\delta g_{ij}=2 k\delta\sqrt{g}$. 
The finite piece in the variation of the  action is then
\eqn\dIfinite{
\delta I_{\rm finite} = \int \left[ \sqrtgo\, \kz^{\,ij}\, 
\delta\!\!\gz_{ij} + \sqrtgz\, \ko^{\,ij}\, \delta\!\!\gz_{ij} 
+ \sqrtgz\,\kz^{\,ij}\,\delta\!\!\go_{ij} -
2\Bigl(\kz~ \delta\!\!\sqrtgo~~+ \ko~\delta\!\!\sqrtgz\,\Bigr) \right]\,.}
This explicitly involves variations of $\go_{ij}$.   These variations 
can be transformed into variations of $\gz_{ij}$ by performing ``integrals by parts''.  
We give the details of one term. Recalling that $\kz_{ij} = \go_{ij}$ the 
third term is 
\eqn\thirdterm{\eqalign{
\sqrtgz~ \; \go^{\, ij} \delta \go_{ij} &=  \sqrt{\gz~}
\; \gz^{\, ik} \gz^{\,jl} \go_{kl}\, \delta\!\!\go_{ij } \cr
&= {1 \over 2} \sqrt{\gz~} \; \gz^{\,ik} 
\gz^{\, jl}\,\delta(\go_{kl} \go_{ij}) \cr
&= -{1 \over 2}\delta \Bigl(\sqrt{\gz~} \; \gz^{\,ik}
\gz^{\,jl}\Bigr) \go_{kl} \go_{ij } + {\rm total \ variation} \cr
&= -{1 \over 2}\delta ( \sqrt{\gz~}) \Tr (\go^{~2}) -
\sqrt{\gz~}\, \delta\!\! \gz^{~ij} (\go^{~2})_{ij} \cr
&= \sqrt{\gz~} \left[ {1 \over 4}\Tr (\go^{~2}) \gz_{ij}
- (\go^{~2})_{ij} \right] \delta\!\!\gz^{~ij}\,.}}
Proceeding in this way, all variations of $\go_{ij}$ can be 
transformed into variations
of $\gz_{ij}$. Up to total variations we finally get
\eqn\dIfinal{
\delta I  = \int\sqrt{\gz~} \left[ \gt_{ij} - {1 \over 8} \gz_{ij}
\Bigl((\Tr \go)^2 - \Tr(\go^{~2}) \Bigr) - {1 \over 2} (\go^{~2})_{ij} 
+ {1 \over 4}\go_{ij} \Tr \go~ \right]  \delta\!\!\gz^{\,ij}\,.}
in full agreement with \dHSS. (Here we have used one equation 
of motion $\Tr(\gt)= {1 \over 4}\Tr \go^{\,2}$ only to make contact with \dHSS.
The above prescription certainly does not require to use 
the solution to the equations of motion.)

\subsec{Chern-Simons gravity in Hamiltonian form and its stress-tensor}

We now apply the above method to Chern-Simons gravity. The Hamiltonian form of ``Lovelock" theories of gravity 
was worked out in \TZ.  As we mentioned before, Chern-Simons gravity is a particular family of theories on which 
all coefficients are correlated.   This has the effect of  enlarging the local symmetry group from $SO(5)$ to 
$SO(1,5)$ (in five Euclidean dimensions).   

To apply our method of stress-tensor renormalization, we write the variation of the action as 
$$
\delta I  = \int \pi^{ij} \delta h_{ij}
$$
where the relation between the momenta $\pi^{ij}$ and the extrinsic curvature for the general Lovelock 
action is \TZ, 
\eqn\pin{ \pi^i_{\ j} = -{1 \over 4} \sqrt{g} \sum_{p\geq 0} \alpha_p \sum_{s=0}^{p-1} C_{s(p)} 
\delta^{[i_1...i_{2s}...i_{2p-1}i]}_{[j_1...j_{2s}...j_{2p-1}j]} \hat R^{j_1j_2}_{\ ~~ i_1 i_2} ...
\hat R^{j_{2s-1} j_{2s}}_{\ ~~ i_{2s-1} i_{2s}} K^{j_{2s+1}}_{\ i_{2s+1}} ... K^{j_{2p-1}}_{\ i_{2p-1}}}
where 
\eqn\csp{ C_{s(p)} = {(-4)^{p-s} \over s![2(p-s)-1]!! }} 
and the hatted tensors refer to $d+1$ dimensional ones (c.f. \GaussCodazi).  

The coefficients $\alpha_p$ depend on the theory under consideration. For 
Chern-Simons gravity they have to be chosen as
\eqn\alphaps{
\a_p={n![2(n-p)]!\over 2^{p-1}(n-p)!}\,.}
In five dimensions the sum in \pin\ contains three terms with coefficients
\eqn\alphafive{
\alpha_{0} = 2\times 4!,  \ \ \ \ \ \ \  \alpha_{1} = 4, \ \ \ \ \ \
\  \alpha_2 = 1\,.}
Inserting them in \pin\ we get\foot{In \TZ\ the signature $-,+,+,...$ was assumed. 
A quick way to transfer the time 
coordinate into $h_{ij}$ is to set $N \rightarrow iN$ (hence $K \rightarrow -i K$) and 
$\sqrt{h} \rightarrow i\sqrt{h}$.}
\eqn\piijone{
\pi^i_{\ j} = \sqrt{h} \left[ 4 \delta^{[ni]}_{[qj]} K^q_n +
\delta^{[nmpi]}_{[q ksj]} \left( -{2 \over
3}K^{q}_{n}K^{k}_{m}K^{s}_{p} +  R^{qk}_{\ \ nm} K^s_p\right)\right]\,.}
Next we write this expression in terms of the FG metric $g_{ij}$ defined as  
$h_{ij} = {1 \over r} g_{ij}$. Using Eq. \Kijdd\ we find 
\eqn\piijtwo{\eqalign{
\pi^i_{\ j} &= {\sqrt{g} \over r^2} \left[ 4 \delta^{[ni]}_{[qj]}
(\delta^q_n -r k^q_n) \right. + \cr  &  \left.  
\delta^{[nmpi]}_{[q ksj]} \left( -{2 \over 3} (\delta^{q}_{n} -
r k^q_n)(\delta^{k}_{m}- rk^k_m)
(\delta^s_p - r k^{s}_{p}) + r R^{qk}_{\ \ nm} (\delta^s_p -r
k^s_p )\right)\right]\,.}}
This expression is much more manageable that it appears. We need to 
look at 
$\pi^{ij}
\delta h_{ij} = \pi^{i}_{\ n} h^{ni} \delta h_{ij} = \pi^{i}_{\ n} 
g^{ni} 
\delta g_{ij}$.
{\it Without making any approximations yet}, we look at the different 
powers of $r$ in
this expression and conclude:

-- The coefficient of $1/r^2$ (the piece containing only Kronecker 
deltas) gives
$\delta (\sqrt{g})$ and hence we discard it.

-- The coefficient of $1/r$ has two contributions. A piece linear in 
 $k$  multiplied by zero! In fact, there is a cancellation between the linear and the cubic terms which, 
of course, happens only for the Chern-Simons action whose coefficients are correlated. There is also a 
piece linear in the curvature. However, it is direct to see that this piece is proportional  to the 
Einstein tensor of the metric $g_{ij}$,  hence it can be written as $\delta (\sqrt{g} R)$, and we 
discard it as well.

-- Finally, the coefficient of $r^0=1$ is
\eqn\dIfinal{
\delta I = -2\int \sqrt{g}\delta^{[nmp i]}_{[q ksj]} \left(
\delta^q_n k^k_{\ m} k^s_{\ p} + {1 \over 2} R^{qk}_{\ nm} k^s_{\ p} 
\right) g^{jn }\delta g_{in}\,.}
Since this term occurs at order zero, its non-zero 
contribution is obtained simply by replacing $g\rightarrow g_{(0)}$ and 
$k\rightarrow g_{(1)}$.  The coefficient is by definition the stress-tensor and it gives exactly 
the component version of \rewrite.

\subsec{The black hole mass}

As a final application of our formula for the stress tensor let us check that it 
provides the correct value of the mass for the Chern-Simons black holes.
Black holes for Chern-Simons gravity exists and have been found in \BTZ. The metric in five dimensions is 
\eqn\metricfive{
ds^2 = -(r^2 - c)dt^2 + {dr^2 \over r^2-c} +r^2 d\Omega_3}
where $c$ is an integration constant related to the mass $M$ in a 
nonlinear way,
\eqn\M{
c = -1 + \sqrt{ 1 +\kappa M}\,,}
and $\kappa$ is a constant that depends on the normalization for Newton's constant. 
This  expression for $M$ was obtained in \BTZ\ using the standard ADM procedure. The minus sign in  
front of the square root provides a solution as well but it is not a black hole. See \BD\ for other 
implications of the ``wrong sign".    

We now put this metric in the FG form.  This is achieved by the simple radial redefinition, 
\eqn\rtorho{r \rightarrow \rho: \ \ \ \ \   r = {1 + \rho  c \over 2 \sqrt{\rho}}}
which brings the metric into the FG form
\eqn\dcbh{
ds^2 = {d\rho^2 \over 4\rho^2} + {1 \over \rho} \left[ {1 \over 4}( 
-dt^2 + d\Omega_3) + 
{c\rho \over 2}(dt^2 + d\Omega_3) + {c^2\rho^2 \over 4} 
(-dt^2+d\Omega_3)\right].}
Note that only three terms in the FG expansion are non zero.  The boundary metric $\gz$ ($\Re\times S_3$) has a 
vanishing Weyl tensor, and hence this is consistent with \KS. The mass, defined as the integral of $T_0$ at the 
sphere at infinity is given by 
\eqn\Mone{
M = \int_{S_3} \tilde T_0 = \int_{S_3} 4\epsilon_{\alpha\beta\gamma} (\Rz^{\alpha\beta} 
+ dx^{\alpha} \wedge \go^{~\beta} )\wedge \go{~^\gamma}}
where we have denoted the coordinates on the sphere by 
$x^\alpha$. 
(Here it is  convenient to  work directly with the 3-form $\tilde T_i$). The curvature on $S_3$ is 
$R^{\alpha\beta} = 4 dx^\alpha \wedge dx^\beta$ and, 
from \dcbh\ we find $\go_{\alpha\beta} =  2c \gz_{\alpha\beta}$, 
which implies $\go^{~ \alpha} =  2c dx^\alpha$. Replacing in the 
formula  for $M$ we find ($\Omega_3$ is the volume of the three-sphere),
\eqn\Mfinal{
M = 4 \Omega_3 (4 + 2c)(2c) = 16 \Omega_3(  2c + c^2)}
which is equivalent to the non-linear relation \M.

Using the general formula \general\ we could also find the mass for a generic theory in any number of 
dimensions. We do not reproduce the result here which has been found using the standard ADM formalism 
in \BTZ (see \OTZ\ for a recent discussion).

\bigskip

\noindent
{\bf Acknowledgments:} S.T. and M.B. would like to thank the Centro de Estudios Cient\'\i ficos in Valdivia, 
Chile and in particular J. Zanelli for hospitality during the initial stages of this work. 
S.T. also acknowledges helpful discussions with M. Hassaine, R. Troncoso and J. Zanelli.  
M.B. was partially supported by FONDECYT (Chile) grants \# 1020832 and \# 7020832, and  
also acknowledges helpful discussions with O. Miscovic and R. Olea.  

\listrefs

\bye